\documentclass{JINST}  
\usepackage{graphicx}
\usepackage{psfig}
\usepackage{epsfig}
\usepackage{subfigure}
\usepackage{amsmath,amssymb}
\usepackage{txfonts}
   \title{Thermal susceptibility of the Planck-LFI receivers}

   \author{
L. Terenzi$^a$\thanks{Corresponding
author.}~, M.J. Salmon, A. Colin$^b$, A. Mennella$^c$, G. Morgante$^a$, M. Tomasi$^c$, P. Battaglia$^d$, M. Lapolla$^d$, M. Bersanelli$^c$, R.C. Butler$^a$, F. Cuttaia$^a$, O. D'Arcangelo$^e$, R. Davis$^f$, C. Franceschet$^d$, S. Galeotta$^g$, A. Gregorio$^{h,g}$, N. Hughes$^i$, P. Jukkala$^i$, D. Kettle$^l$, M. Laaninen$^k$, P. Leutenegger$^d$, R. Leonardi$^m$, N. Mandolesi$^a$, M. Maris$^g$, P. Meinhold$^m$, M. Miccolis$^d$, N. Roddis$^f$, L. Sambo$^a$, M. Sandri$^a$, R. Silvestri$^d$, J. Tuovinen$^j$, L. Valenziano$^a$, J. Varis$^j$, F. Villa$^a$, A. Wilkinson$^f$, A. Zonca$^n$\\
\llap{$^a$}INAF - IASF Bologna, 
	via P. Gobetti, 101 -- I40129 Bologna, Italy\\
\llap{$^b$}Instituto de Física de Cantabria (CSIC-UC), 
	Av. Los Castros s/n. 39005 Santander-Spain\\
\llap{$^c$}Universit\'a degli Studi di Milano, 
	Via Celoria 16, 20133 Milano, Italy\\
\llap{$^d$}Thales Alenia Space Italia, Sede di Milano, 
	S.S. Padana Superiore, 290 -- I20090 Vimodrone, Italy\\
\llap{$^e$}Istituto di Fisica del Plasma CNR, 
	via Cozzi 53, 20125 Milan, Italy\\
\llap{$^f$}Jodrell Bank Centre for Astrophysics, 
	Alan Turing Building, The University of Manchester, Manchester, M13 9PL, UK\\
\llap{$^g$}INAF Osservatorio Astronomico di Trieste, 
	via Tiepolo, 11 Trieste, I-34143, Italy\\
\llap{$^h$}University of Trieste, Department of Physics, 
	via Valerio, 2 Trieste I-34127, Italy\\
\llap{$^i$}DA-Design Oy, 
	Keskuskatu 29, FI-31600, Jokioinen, Finland\\
\llap{$^j$}MilliLab, VTT Technical Research Centre of Finland, 
	Espoo, Finland\\
\llap{$^k$}Ylinen Electronics Oy, 
	Kauniainen, Finland\\
\llap{$^l$}School of Electrical and Electronic Engineering, The University of Manchester, 
	Manchester, M60 1QD, UK\\
\llap{$^m$}Department of Physics, University of California, 
	Santa Barbara, CA 93106, USA\\
\llap{$^n$}INAF IASF Milano, 
	Via Bassini, 15, 20133, Milano, Italy\\
	Email: \email{terenzi@iasfbo.inaf.it}
}

  \abstract
   {This paper describes the impact of the Planck Low Frequency Instrument front end physical temperature fluctuations on the output signal.
   The origin of thermal instabilities in the instrument are discussed, and an analytical model of their propagation and impact on the receivers signal is described. The experimental test setup dedicated to evaluate these effects during the instrument gound calibration is reported together with data analysis methods.
Finally, main results obtained are discussed and compared to the requirements.}


\begin{document}
%

\section{Introduction}

Planck is an ESA mission designed to map with high precision the 
angular distribution of the Cosmic Microwave Background (CMB). The 
Planck observations are expected to produce major steps forward for 
precision cosmology as well as for Galactic and extragalactic 
millimeter-wave astrophysics \cite{tauber}.\\ 
The Planck measurements will span over a wide range of frequencies by means of the two instruments on board: the Low Frequency Instrument (LFI, \cite{mandolesi}) ranging from 30 to 70 GHz, and the High Frequency Instrument (HFI, \cite{lamarre}), from 100 to 850 GHz.\\
LFI is an array of 22 pseudo-correlation radiometers (two for each feed horn, \cite{LFI}), whose core elements are HEMT-based low noise amplifiers located in the Front End Modules (FEMs). Radiometers are labelled as Main and Side arm for each of the Radiometer Chain Assemblies (RCAs) fed by one horn. The signal coming from the sky is compared continously to the emission of the 4 K reference load (4K RL, \cite{valenzia}), consisting of small blackbodies connected to the High Frequency Instrument shield at a temperature of about 4 K.\\
The ambitious Planck scientific goal to finely map sky temperature differences at level of $\mu K$ requires a strict control of time variations in the signal due to environmental systematic effects such as thermal or electrical instabilities.\\
In this paper the susceptibility of the radiometers to temperature fluctuations is studied.\\
The presence of active coolers, mainly the 20 K sorption cooler \cite{morgante_SC}, in the satellite cryogenic chain produces temperature instabilities in different stages of the instruments.\\
The two sorption cooler cold ends give the reference temperature to the LFI focal plane and main frame and serve as precooling stage for the 4 K cooler of the HFI.\\
Fluctuations generated in the sorption cooler are thus responsible for temperature oscillations both in the LFI amplifiers, located on the focal plane, and in the 4K Reference Load, mounted on the HFI outer shield.\\ 
These temperature fluctuations produce a signal variation which mimics the observed CMB temperature anisotropies. The accurate knowledge of the impact of physical temperature fluctuations on output antenna temperature fluctuations is therefore fundamental to estimate the error induced by this systematic effect on the LFI observations.\\
In this paper we focus on the effect of temperature fluctuations in the instrument front end. They have been modeled and subsequently measured during two instrument ground test campaigns: 

\begin{itemize}
 \item the Radiometer Chain Assembly (RCA) level test \cite{villa_RCA}, where each single radiometer chain associated to the same sky horn was tested indipendently 
 \item the Radiometer Array Assembly (RAA) level test \cite{mennella_RAA}, where all the RCAs were integrated together and the whole Low Frequency Instrument was tested
\end{itemize}
The theoretical fundamentals of LFI radiometer susceptibility to thermal effects are outlined in Section 2.\\ 
In Section 3, we discuss in detail the susceptibility measurements, describing the experimental setup and data analysis methods and summarizing the obtained results.\\
Finally in Section 4, the main conclusions are drawn.

\section{Effect of focal plane thermal fluctuations on LFI radiometers signal}

\subsection{The thermal systematic effects}

The active elements of the Planck LFI radiometers are located in the Front End Modules (FEMs) and Back End Modules (BEMs).\\
Fluctuations in the physical temperature of these modules affect the basic properties of the low noise amplifiers, such as noise temperature and gain, causing a correlated fluctuation in the output signal.\\
An analytical expression for the measured radiometer output, $p_{out}$, can be expressed as a function of the sky and reference load input signals, and of parameters which are temperature dependent, such as amplifiers' gain and noise temperature, which we indicate here generically as $K_i$, \\ $p_{out}\equiv p_{out}(T_s,T_{4K},K_i)$ (see \cite{seiffert} for details).\\ 
The sky signal fluctuation, $\delta T_s$, equivalent to the systematic effect induced by thermal fluctuation, $\delta T_{phys}$, can then be evaluated from:

\begin{equation}
	\frac{\partial{p}}{\partial{T_s}} \cdot \delta T_s = \frac{\partial{p}}{\partial{T_{phys}}} \cdot \delta T_{phys} 
\end{equation}
so that we can define a radiometric transfer function for physical temperature fluctuation:

\begin{equation}
  T_f^{id} = \frac{\frac{\partial{p}}{\partial{T_{phys}^{id}}}}{\frac{\partial{p}}{\partial{T_s}}}
\label{tf_eq}
\end{equation}
where the index $id$ identifies the different temperature stages whose fluctuations have an impact on the signal output, FEM, BEM, 4 K RL being the most relevant.\\
Generally the dependence of $p_{out}$ on temperature is not explicit, but embedded in the temperature dependence of the various instrument parameters $K_i$ , so that we have:

\begin{equation}
	\frac{\partial{p}}{\partial{T_{phys}}} \equiv \sum_i \frac{\partial{p_{out}}}{\partial{K_i}} \cdot
	\frac{\partial{K_i}}{\partial{T_{phys}}}
\end{equation}

Once the transfer function $T_f$ is defined, we can characterize how temperature changes in the instrument impact on the signal by means of the relation:

\begin{equation}
 \delta T_s = T_f \cdot \delta T_{phys} 
\label{slope_eq}
\end{equation}

As thermal instabilities mimic signal variations in a direct way, in defining the requirement levels for this kind of systematic effect, we have to treat carefully the spin synchronous fluctuations which would be undistinguishable from the sky signal.\\
The maximum allowable spurious signal caused by focal plane thermal fluctuations must be lower than 1 $\mu$K (0.9 $\mu$K for a generic periodic fluctuation and 0.45 $\mu$K for a spin synchronous fluctuation, \cite{LFI}).\\ 
As explained in the following and finally discussed in the last section, different properties of the instrument have to be taken into account, in order to accurately control this systematic effect and verify the compliance with requirements. 
 
\subsubsection{The source of front end thermal instability: the hydrogen sorption cooler}

\begin{figure}[h!]
	\centerline{\epsfig{file=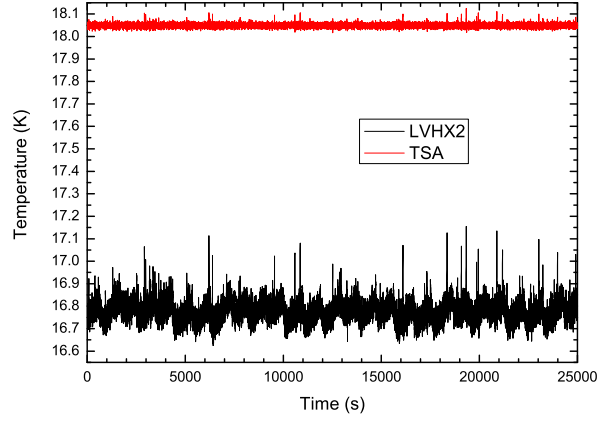,width=9.cm}}
	\caption{A typical temperature curve of the LVHX2 sorption cooler cold end (black line) during ground tests. The temperature after thermal stabilization is also shown (red line).}
\label{sc_plot}
\end{figure}

The LFI first stage of signal amplification is located in the Focal Plane Unit (FPU) of the instrument, cooled at 20 K by the hydrogen sorption cooler \cite{morgante_SC}.\\
The cooler consists of six compressor elements absorbing and desorbing hydrogen gas in a sequential way to create liquid in its cold ends (liquid-vapour heat exchangers, LVHXs) through a JT expansion. The temperature at the cooler cold ends fluctuates at the level of $\sim$ 400 mK peak to peak, with a spectrum dominated by two main frequencies: the compressor element cycle, typically in the range 600--1000 s, and at the period of the whole cooler, six times longer than the previous one.\\
In order to significantly reduce undesired fluctuations, an active temperature control stage, the Thermal Stabilization Assembly (TSA), based on a PID algorithm was inserted between the LVHX2 and the LFI main frame.\\
Typical sorption cooler temeprature fluctuations are shown in Fig. \ref{sc_plot}, which displays two timestreams: the black line corresponds to the cold end temperature and the red line corresponds to the temperature downstream of the TSA, whose effect is to decrease the peak--to--peak amplitude to below 100 mK; in particular, it is effective in the lower frequencies part of the spectrum and the two main periods of the cooler are strongly reduced.

Temperature fluctuations propagate through the LFI mechanical structure and are damped by the LFI front end thermal mass, which acts as a low pass filter. This effect was first studied by means of a dedicated thermal model of the instrument and then measured during instrument level ground test (see \cite{tomasi_TM} for details).

\subsection{Transfer function between front end temperature variations and the radiometer signal}

If $\delta T_{phys}^{FEM}$ is the temperature fluctuation at the level of the front-end module, then the systematic variation induced in antenna temperature ouput is given by:\\
$\delta T_s = T_f^{FEM} \cdot \delta T_{phys}^{FEM}$ \\
The main contributions to the radiometer output equation coming from the instrument 20 K stage are given by the amplifiers gain and noise temperature, and the attenuation of the signal coming from the sky, the feed horns and OMT insertion losses, and from the reference load, by the reference horn loss. The properties related to the electronic active devices are known to be affected at a significant and measurable level by fluctuations of the order of a few K, while passive losses, which depend on geometrical and material properties, are considered constant in the typical temperature range where the fluctuations occurred.\\
Applying Eq. \ref{tf_eq} to the radiometer differenced output expression (reported in the appendix, Eq. \ref{full_out}), an analytical expression for the transfer function can be found as:

\scriptsize

\begin{multline}
	T_f^{FEM} = \left(\frac{L_{fh-OMT}}{(G_{F1}+G_{F2}) \cdot (1-r) + 2 \cdot \sqrt{G_{F1} \cdot G_{F2}} \cdot (1+r) } \right) 
							\cdot [\left(\frac{\partial G_{F1}}{\partial T_{phys}^{FEM}}\right)\cdot\left((T_{sky} + T_{4K}+2\cdot T_{nF1})
							\cdot (1-r) + \sqrt{\frac{G_{F2}}{G_{F1}}}\cdot (1+r)\right) + \\
							\left(\frac{\partial G_{F2}}{\partial T_{phys}^{FEM}}\right)\cdot\left((T_{sky} + T_{4K} + 2 \cdot T_{nF2}) 
							+ \cdot (1-r) + \sqrt{\frac{G_{F1}}{G_{F2}}} \cdot (1+r)\right) +  
							\left(\frac{\partial T_{nF1}}{\partial T_{phys}^{FEM}}\right) \cdot (2\cdot G_{F1}\cdot (1-r))
							+ \left(\frac{\partial T_{nF2}}{\partial T_{phys}^{FEM}}\right) \cdot (2\cdot G_{F2}\cdot (1-r)) + \\
							\left(1-\frac{1}{L_{fh-OMT}}\right)\cdot\left((G_{F1}+G_{F2})\cdot (1-r)+2\cdot \sqrt{G_{F1} \cdot G_{F2}} \cdot (1+r) \right) + 
							\left(1-\frac{1}{L_{4K}}\right)\cdot\left( (G_{F1}+G_{F2})\cdot (1-r)-2\cdot \sqrt{G_{F1} \cdot G_{F2}} \cdot (1+r) \right)]
\label{fem_tf}
\end{multline}

\normalsize

where:
\begin{itemize}
	\item $L_i$ are insertion losses either for the feed horn--OMT system or for the 4K horn antenna; in our analysis we assume their values estimated from measurements at room temperature;
	\item $r$ is the {\it gain modulation factor} used to balance the sky and reference output signals; its value is evaluated from the ratio of sky to reference channel mean voltage values; 
	\item $G_{Fi}$ are the front end amplifier gains, whose typical value is about 35 dB  
	\item $T_{nFi}$ are the front end amplifier noise temperatures, evaluated from dedicated tests
\end{itemize}

The insertion losses, noise temperatures and gains are estimated from dedicated ground measurements, while the physical temperatures, sky temperature, and the $r$ parameter are quantities measurable in flight, it is thus apparent from the above expression that our ability to predict the transfer function and thermal susceptibility in flight depend fundamentally on our knowledge of $\frac{\partial G_{F}}{\partial T_{phys}^{FEM}}$ and $\frac{\partial T_{nF}}{\partial T_{phys}^{FEM}}$.\\ 
A requirement of our analysis is then to estimate these important parameters. 

\section{Thermal susceptibility test}

\begin{figure}[ht!]
	\centerline{\epsfig{file=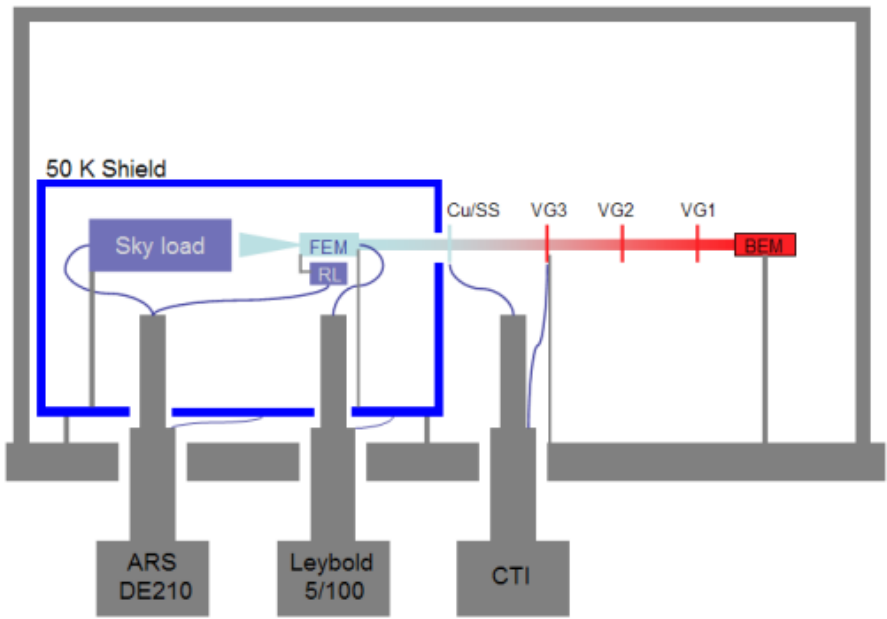,width=12.cm}}
	\caption{A schematic view of the 30 and 44 GHz RCA thermal setup integrated in the cryogenic chamber (from \cite{terenzi_RCA}). The feed horn receives the signal from the sky simulator in front of it; the signal is mixed in the FEM with the signal coming from the reference load and amplified. The long waveguides finally transmit the signal to the BEM detectors. In the thermal susceptibility tests, the temperature of the FEM is varied while reference and sky load temperatures are kept constant.}
\label{RCA_Scheme}
\end{figure}

During the flight model ground test campaign, the thermal susceptibility of front end modules to temperature fluctuations was measured both at the level of single RCAs and at the level of integrated instrument in the RAA cryogenic facility.\\ 
The RCA tests were performed with environmental and boundary conditions under a better control because of smaller dimensions involved and a more accurate monitoring was achieved due to dedicated temperature sensors in a smaller number of relevant interfaces; reference results are then taken from these.\\ 
In Fig. \ref{RCA_Scheme}, a schematic view of a single RCA integrated in the test chamber \cite{terenzi_RCA} is shown.

The RAA thermal susceptibility tests have been performed in order to check consistency with the RCA results. Unfortunately, they were performed in non-ideal conditions so that a straight comparison with RCA tests is actually quite poor.\\

\subsection{Test methods and procedures}

\begin{figure}[!h]
\subfigure[]{\epsfig{file=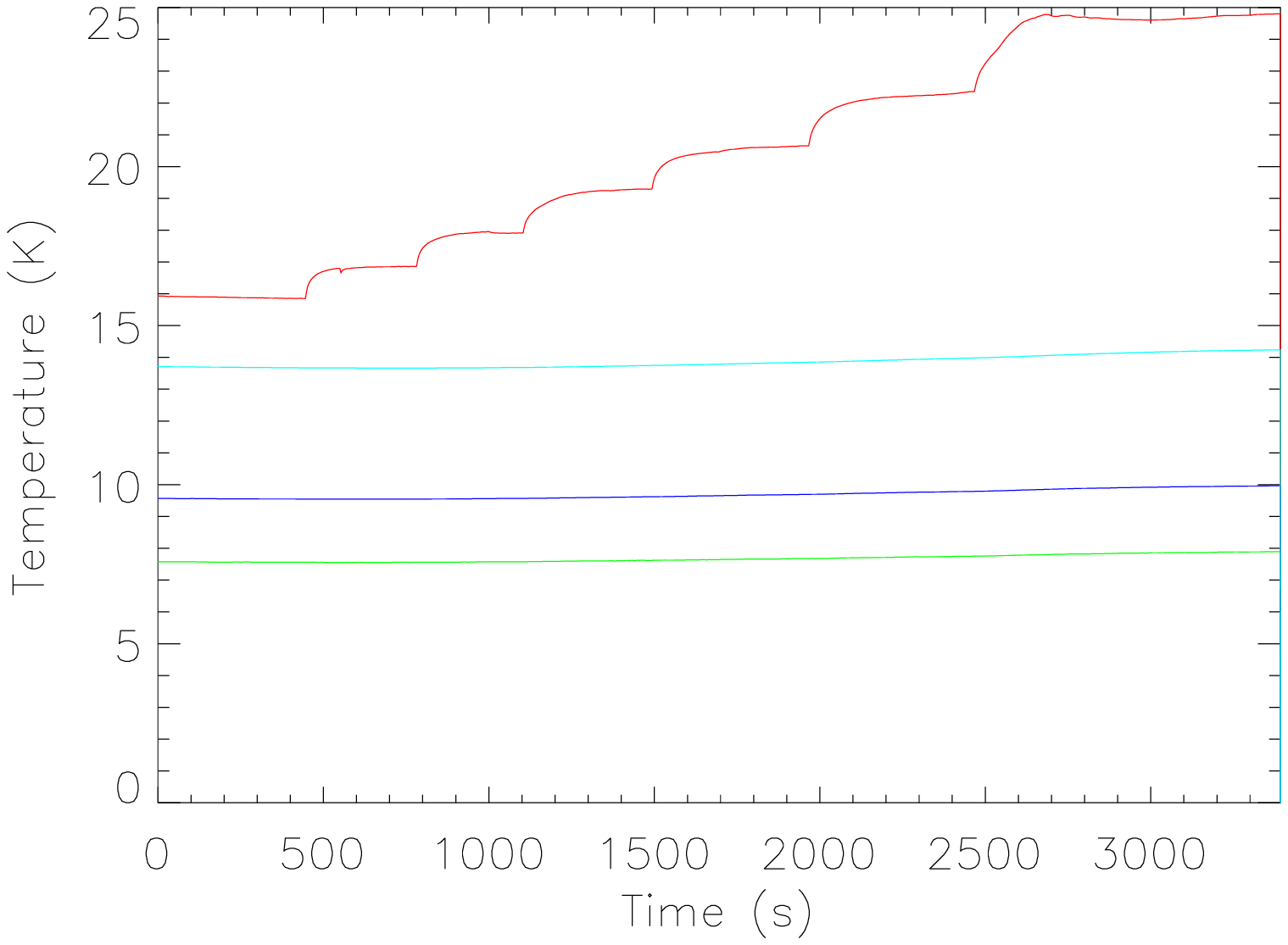,width=7.cm}}
\subfigure[]{\epsfig{file=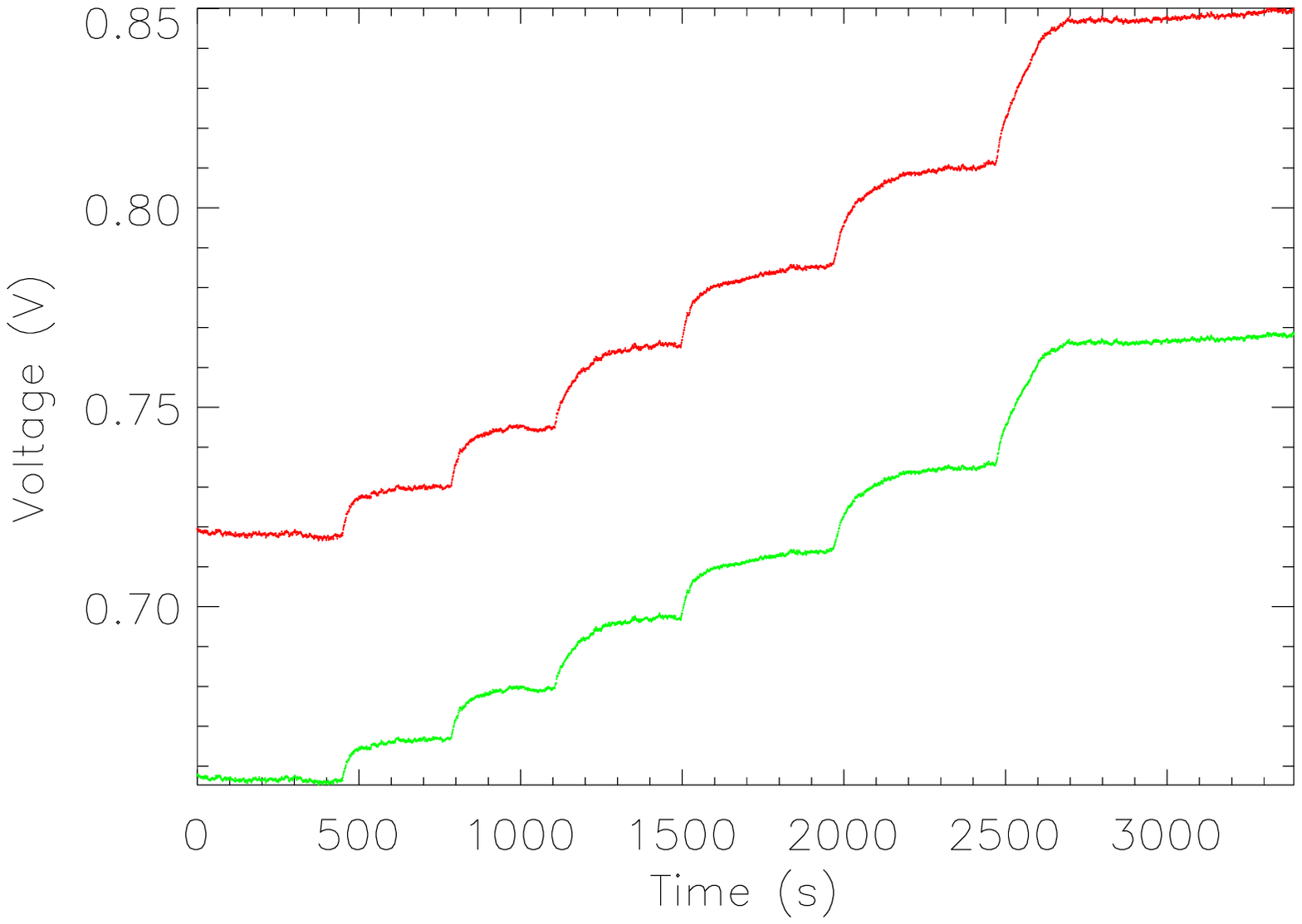,width=7.cm}}
\caption{An example of THF test. Data are taken from RCA22 test, one of the 70 GHz radiometers.
In (a) temperature data for FEM (red), ref (green), sky backplate (blue) and absorber (light blue) are shown: as evident here the FEM temperature has to change significantly, while reference and sky loads have to be kept at a temperature as stable as possible. Corresponding detector outputs are displayed in (b): sky is in red and ref in green.}
\label{rca_tv}
\end{figure}

The test philosphy and, accordingly, its procedure are very simple.
The test was performed by changing the temperature of a single FEM (RCA test), or of the whole Focal Plane Unit (RAA test), and by measuring the radiometer output during the subsequent steady state period; the number of the steps was chosen in order to scan a temperature range of about 5K around the nominal value.
Typical temperature curves and voltage output obtained during RCA tests are shown in Fig. \ref{rca_tv}.\\
In order to disentangle variations in the signal due to changes of FPU temperature from other sources of instability, the temperatures of all the main parts of the cryochamber, in particular sky and reference loads, are monitored and their stability is controlled at the level of few mK.

\subsection{Experimental setup}

\begin{figure}
	\centerline{\epsfig{file=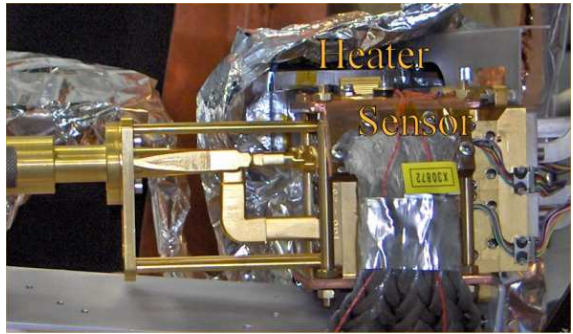,width=7.5cm}}
	\caption{In the 30/44 GHz setup the heater and the sensor dedicated to the control of the FEM temperature are located on a flange directly 
	screwed to the front end.}
	\label{rca_fem}
\end{figure}

The RCA and RAA cryogenic chambers are described in \cite{terenzi_RCA} and \cite{morgante_RAA}, respectively.\\ 
The chamber thermal environment reproduces accurately the actual flight environment, with the exception of the sky and reference loads temperatures.\\
However, the most important test features are the temperature stability of the two loads, which are kept within an optimal level, so that the test results, obtained in a optimized and controlled environment, can be easily extrapolated to the flight conditions. This is also true for al of the test described here.\\
The temperature control for the FEM thermal susceptibility tests (THF tests) is implemented through heaters, mounted on the interface between the front end and the flanges connected to the chamber reference cold end at 20 K, and temperature sensors, used to control the heaters and to monitor the FPU temperatures.\\
In the case of the RCA tests, a single FEM is connected to a cold finger with a copper flange where one heater and one sensor are mounted (Fig. \ref{rca_fem}).

\begin{figure}
	\centerline{\epsfig{file=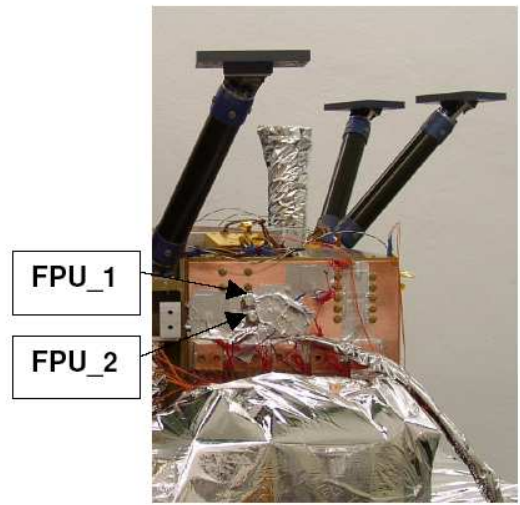,width=8.0cm}}
	\caption{Main frame and FPU setup for the thermal vacuum tests. Nominal and redundant sensor dedicated to the control of the FPU temperature are
	shown, while heaters are fixed by Aluminum tape.}
	\label{raa_fpu}
\end{figure}

\begin{figure}
	\centerline{\epsfig{file=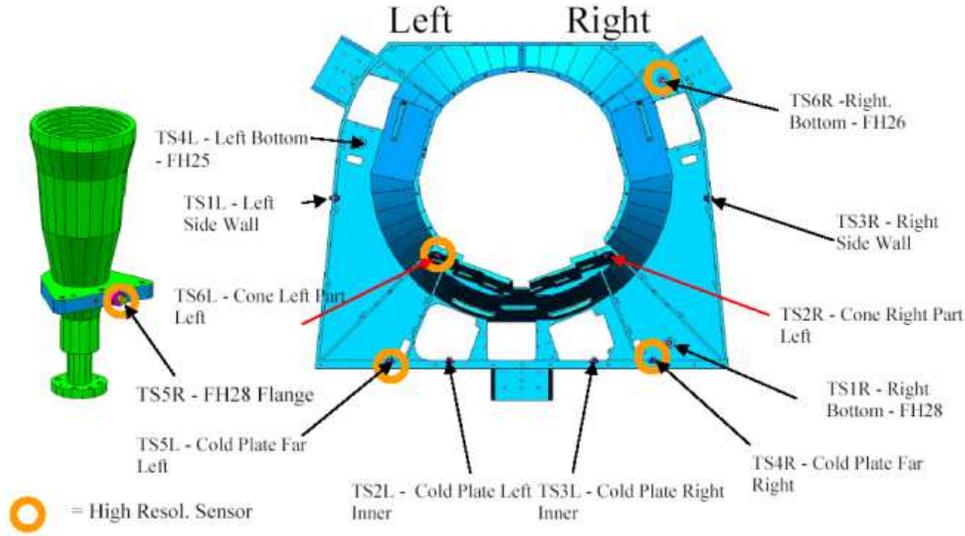,width=14.cm}}
	\caption{Sensors location in the LFI FPU and main frame. High sensitivity sensors are marked with circles.}
	\label{sensor_map}
\end{figure}

At instrument level, the heater and corresponding sensor are mounted on a large copper flange, which mimics the sorption cooler cold end interface to the LFI main frame (Fig. \ref{raa_fpu}). The corresponding temperature at the level of single FEMs is chosen from one of the sensors, whose detailed positions are shown in Fig. \ref{sensor_map}. 

\subsection{Data analysis} 

\begin{figure}[!here]
\centerline{\epsfig{file=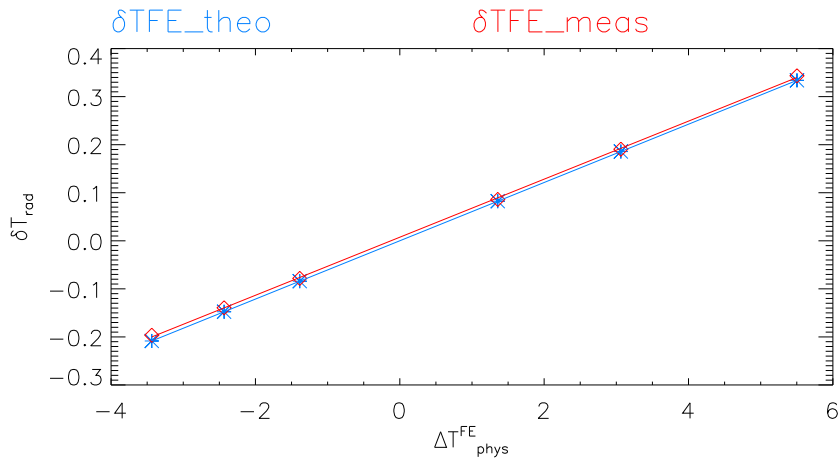,width=12.cm}}
\caption{Radiometer susceptibility analysis. Data are taken from the RCA22 test again. 
Experimental measured (red) vs. Analytical prediction (blue). Measurement errors are within the symbol sizes.}
\label{rana_out}
\end{figure}

For each value of the front-end temperature, $T_{phys}^{FE}$, we measured the receiver response (see Fig. \ref{rca_tv}), and then calculated the average receiver differential output, $\delta T_{ant}$, using the same value of the gain modulation factor, $r$, that was calculated in nominal front-end temperature conditions.\\
It is then possible to produce a plot of $\delta T_{ant}$ vs. $\delta T_{phys}^{FE}$, where $\delta T_{phys}^{FE} = T_{phys}^{FE} - T_{phys}^{FE}$(nominal) (see Fig. \ref{rana_out}).\\
From Eq. \ref{slope_eq}, the transfer function is the slope of the curve obtained by linear fitting the points in the plot.\\
As explained above, the main unknown susceptibility parameters in Eq. \ref{fem_tf} are the amplifier noise temperature and gain variations with temperature.\\
In order to estimate them, their values is varied in the analytical expression of Eq. \ref{fem_tf}, so to obtain a value consistent with the 
measured one. The predicted points and curve are in blue in Fig. \ref{rana_out}.

\begin{figure}[!h]
\centerline{\epsfig{file=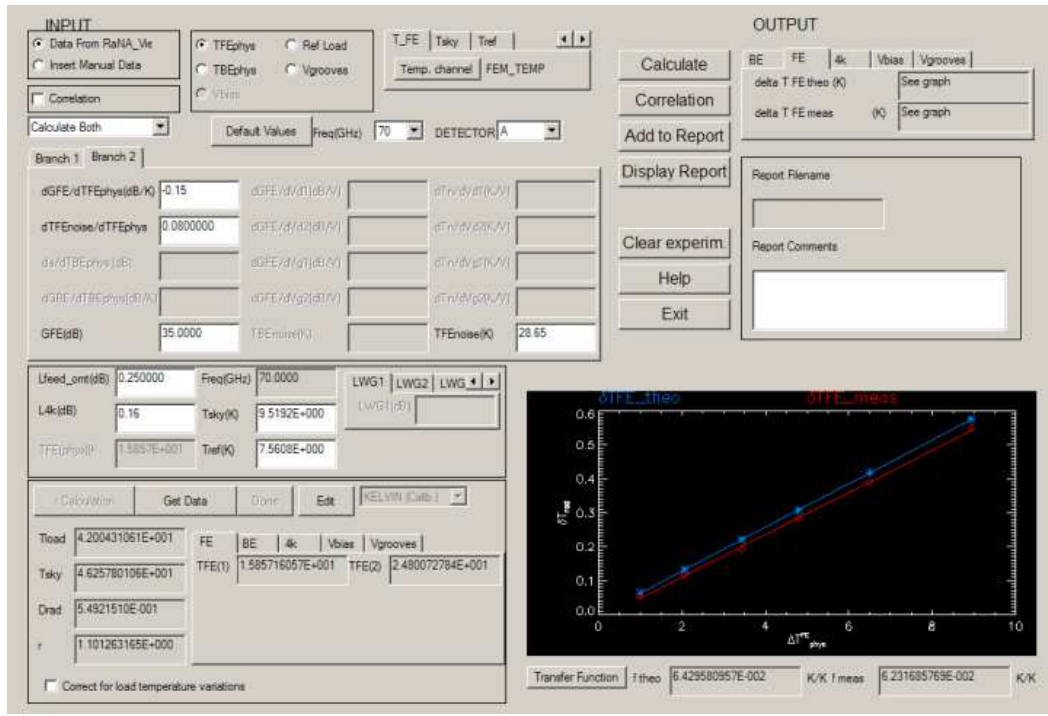,width=14.cm}}
\caption{Layout of the RaNA\_Susc module of the LIFE data analysis tool. It allows to get automatically voltage and temperature data useful for the estimation of the thermal susceptibility, calculate the transfer function and compare it to the analytical one, using default or user-defined parameters.}
\label{rana_susc}
\end{figure}

The RaNA data analysis tool (\cite{tomasi_LIFE}) for the RCA tests has a built-in THF module in order to automatically select useful data for this kind of test, calculate the fitted and measured values, and write a report of the test with table of results, best fit parameters and plots (Fig. \ref{rana_susc}).

These results are our best guesses for the intrinsic properties of the LFI amplifiers affected by the temperature fluctuations. The analytical values  obtained assume a symmetric behaviour of the amplifiers corresponding to the two legs of the radiometer analyzed.

\section{Results and discussion}

\begin{figure}
	\centerline{\epsfig{file=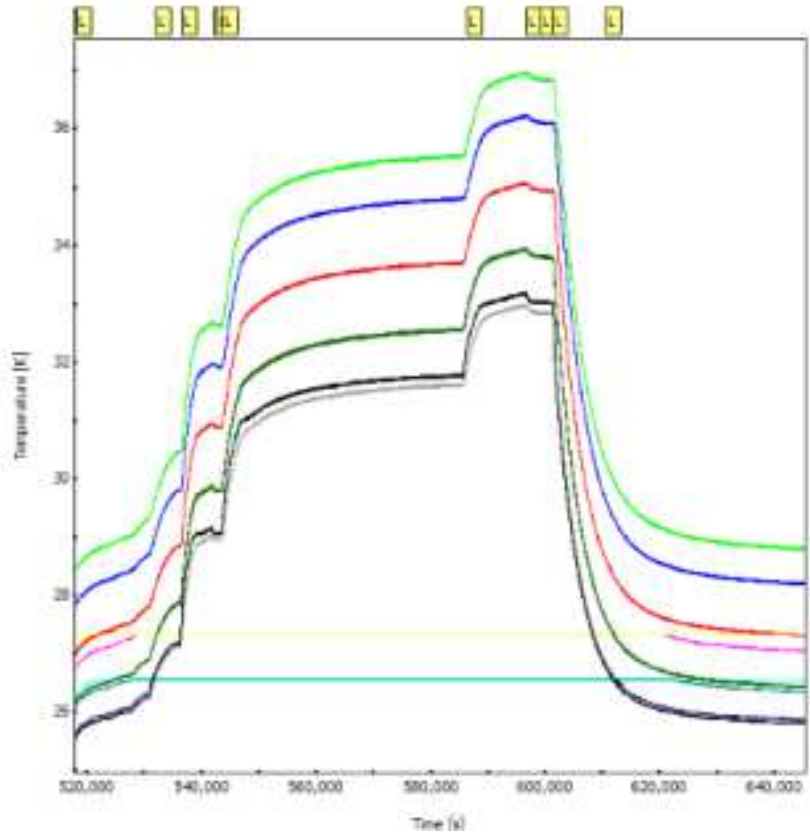,width=11.cm}}
	\caption{FPU sensors temperature curve. The five high resolution sensors are out of range for most of the steps. This is evident from the straight horizontal lines in the middle of the plot: when sensors go higher than their upper calibration limit their output is fixed to that value until their temperature goes back in the calibration curve range.}
	\label{raa_sensors}
\end{figure}

The RCA calibration tests produced the most accurate results due to a dedicated test setup.\\
Actually, some problems during instrument level verification tests did not allowed a straight comparison between test results. They were due to radiometer settings (non optimal bias set during the first run) or to large thermal drifts (as occurred in the second run).\\
Moreover, not all the radiometer amplifiers are closely monitored by thermometers (Fig. \ref{sensor_map}) and the choice of the best temperature sensor to associate to the FEM under test was not always straightforward, and finally the most accurate sensors have a limited range of calibration. 
Since the temperatures reached during the instrument level thermal susceptibility test were outside this calibration range, a number of thermometers were therefore not used for the data analysis (see Fig. \ref{raa_sensors}).\\

\begin{figure}[h!]
 \begin{tabular}{c c}
	\epsfig{file=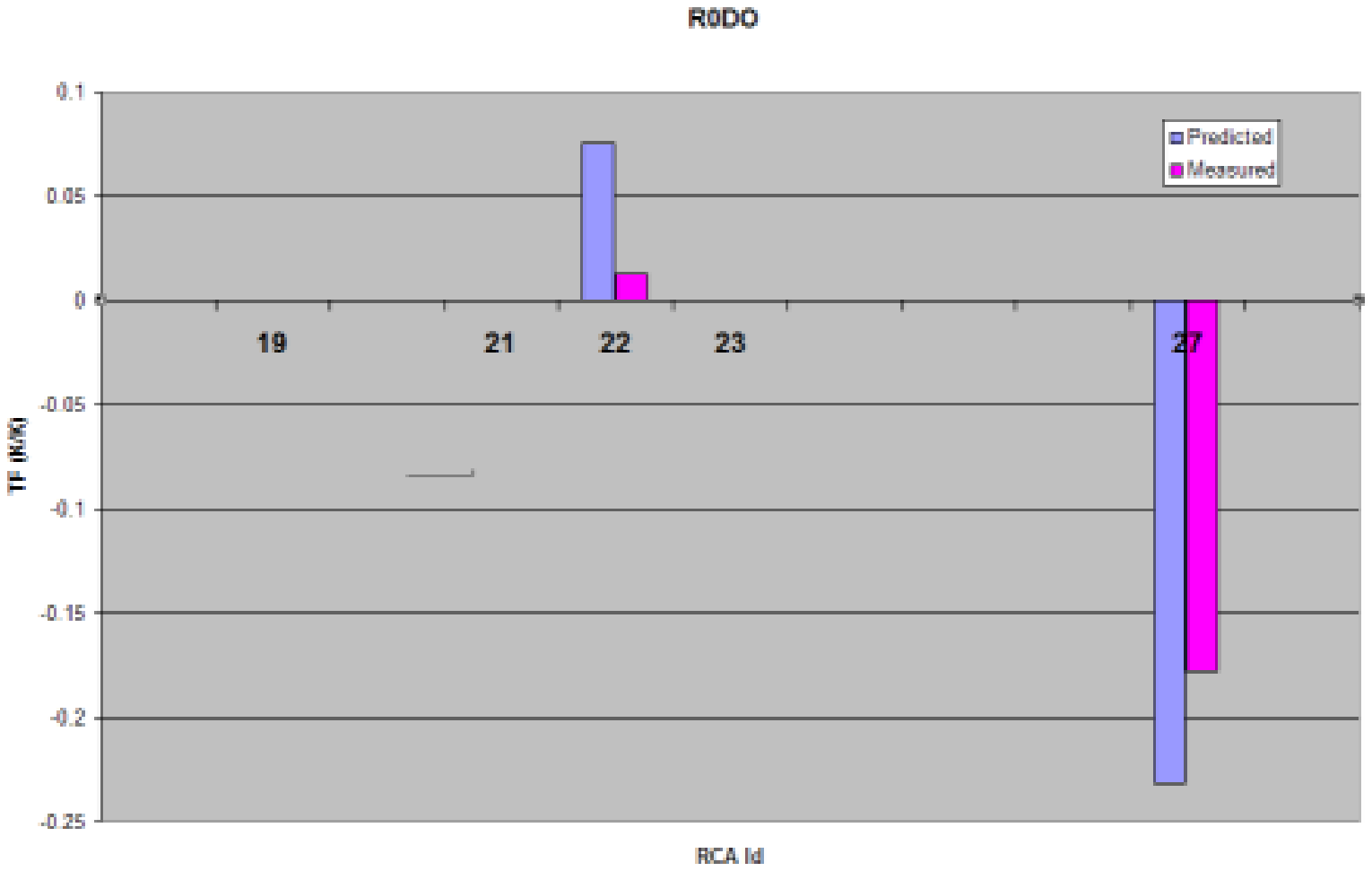,width=7cm} & \epsfig{file=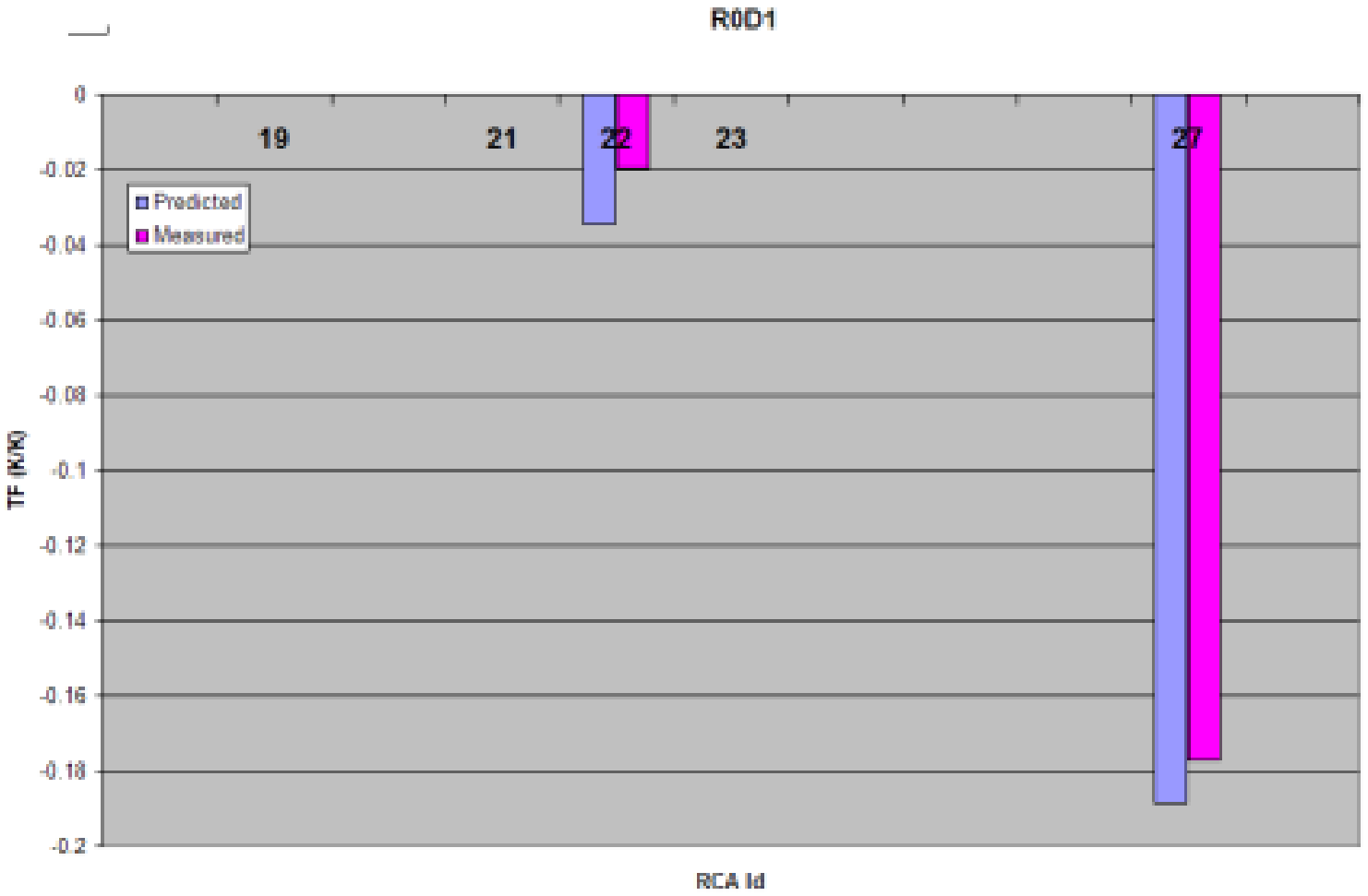,width=7cm} \\
	\epsfig{file=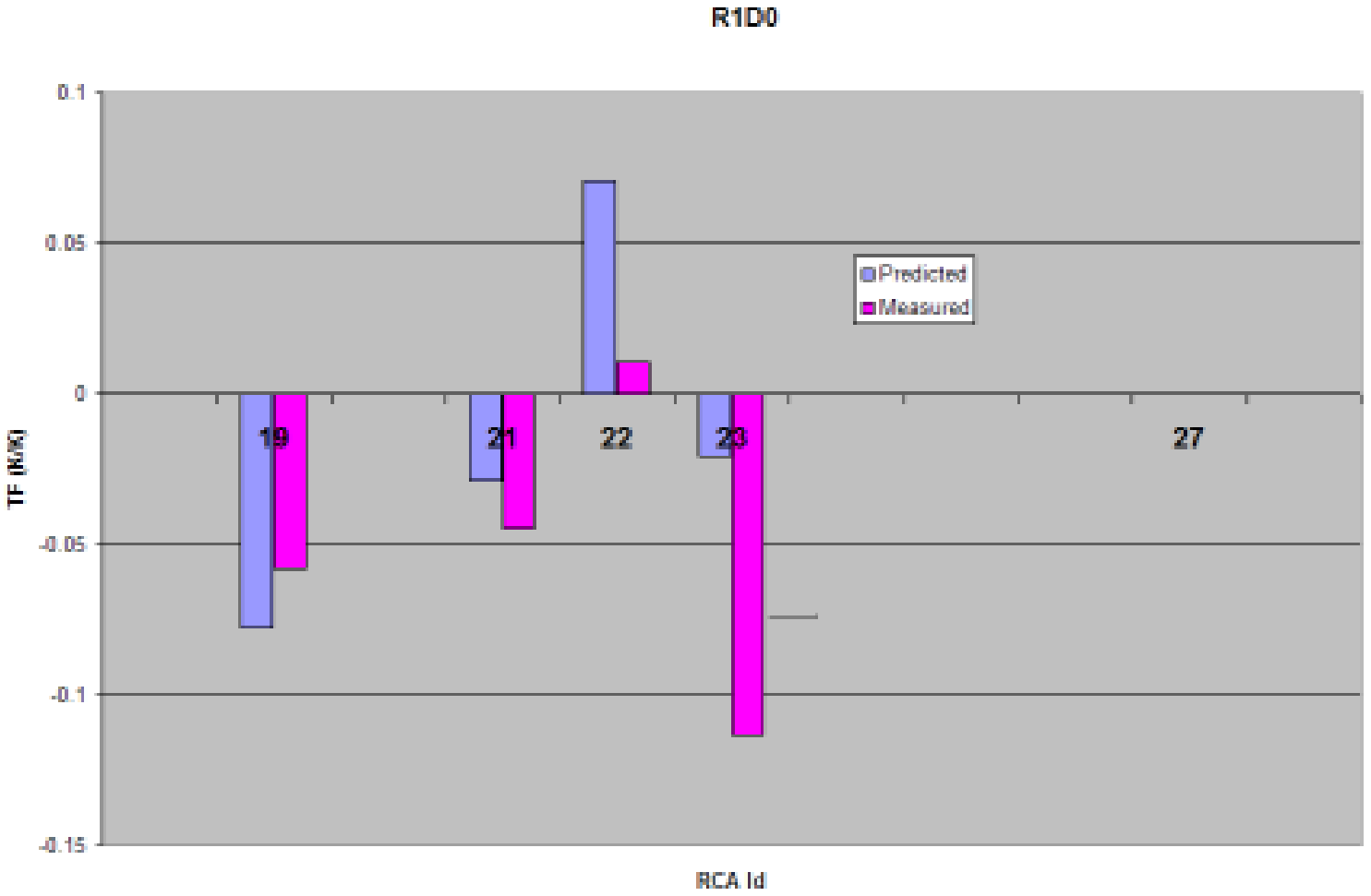,width=7cm} & \epsfig{file=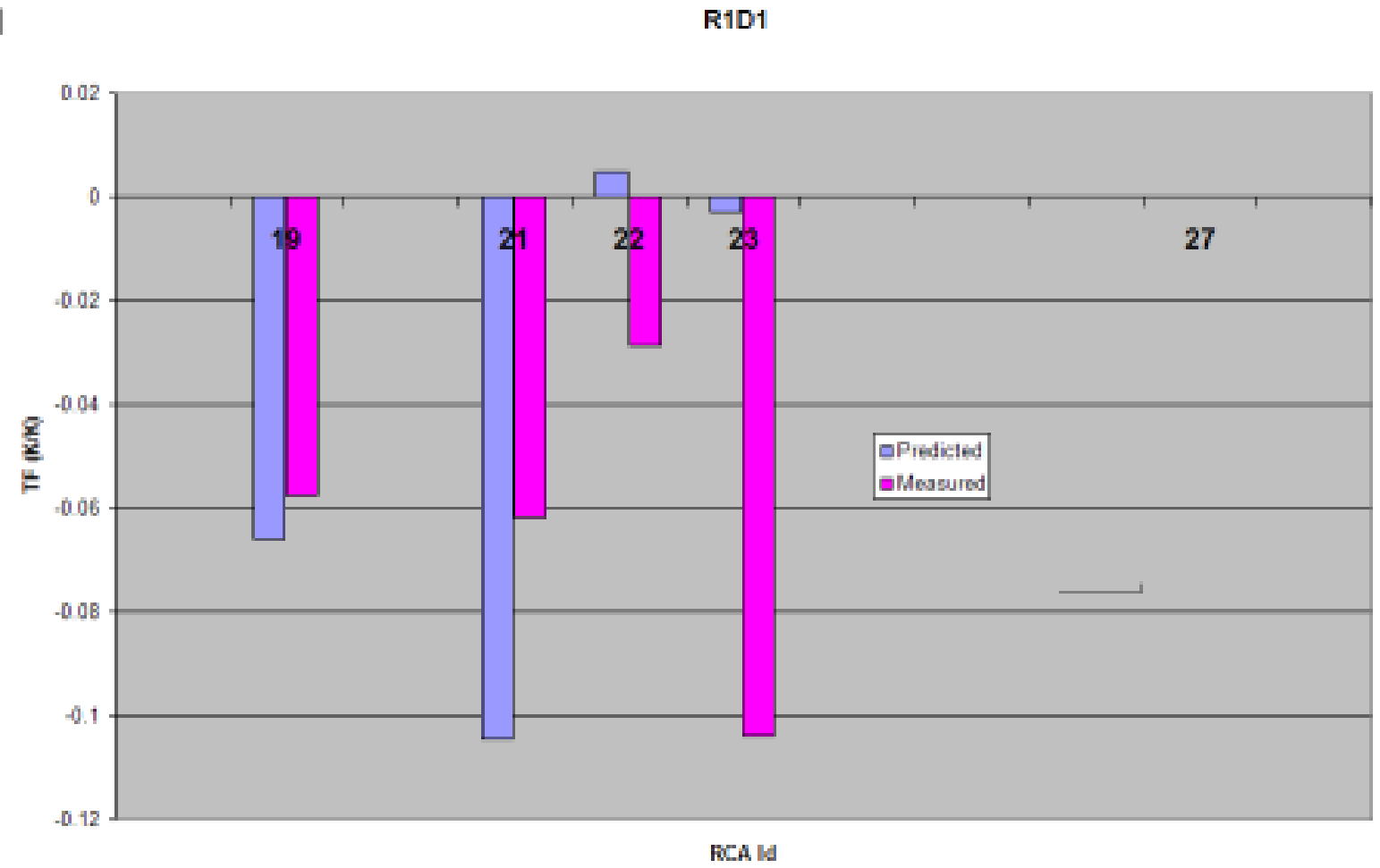,width=7cm} \\
 \end{tabular}
\caption{Comparison between predicted and measured transfer functions during instrument level tests for the channels properly biased. Radiometers Main and Side are labelled as R0 and R1, while detectors outputs as D0 and D1, for each radiometer leg. The predicted transfer functions are obtained from the analytical expression using radiometric parameters found in RCA tests and environmental parameters measured in the RAA test itself.}
\label{raa_vs}
\end{figure}

Fig. \ref{raa_vs}, shows a comparison between the predicted transfer functions and the ones measured during RAA tests.
The predicted results comes from the analytical formula \ref{fem_tf}, using RAA measured environmental data and radiometer intrinsic properties ($\frac{\partial G_{F}}{\partial T_{phys}^{FEM}}$ and $\frac{\partial T_{nF}}{\partial T_{phys}^{FEM}}$) estimated from RCA tests. This is reported only for those radiometer channels correctly biased during the RAA test performed in more stable conditions.\\ 
Considering the limitations affecting the measurement the agreement is good for most of the channels.\\
Detailed results from the RCA flight model test campaign are reported in the Appendix.\\
All values are contained in the range (1 $\div$ 100) $\frac{mK}{K}$, with higher values (50 to 120 $\frac{mK}{K}$) for the 70 GHz channels (except RCA 21 Side radiometer), so that depending on the radiometer, the physical temperature fluctuations of the front end module are reduced by a factor of 10 to 1000. This effect is actually flat in the frequency space and it is applied at the end of the path of thermal instabilities already filtered by the thermo-mechanical structure of the LFI focal plane (detailed in \cite{tomasi_TM}).\\
Taking into account the source of temperature fluctuations, reduced by means of the active control at the order of 100 mK peak--to--peak and 
considering all the reduction factors, we can estimate to less than 1 mK peak--to--peak in the radiometer output data stream.
A further reduction of the effect at the different frequency components of the order of about 200 - 500 times occurs in the map making and destriping procedures, as estimated in \cite{mennella_2001}.\\
The final result is that temperature fluctuations generated by the 20 K sorption cooler and propagated through the LFI focal plane are kept at the level of the required error budget.\\
A by-product of our analysis is the information about the temperature dependance of amplifiers noise temperature and gain, which are shown in the Appendix tables. The range of gain susceptibility is -0.01 -- -0.08 $\frac{dB}{K}$ with some major exception for the RCA 22 and 28.
The temperature variations of amplifiers noise temperatures ranges between 0.1 and 0.8 $\frac{K}{K}$.\\
Due to dependance of the transfer functions on input temperatures and $r$ parameter (Eq, \ref{fem_tf}), these are important sources of information allowing the estimation of the impact of in-flight temperature fluctuations on the measured signal and on CMB recovery.

\acknowledgments
Planck is a project of the European Space Agency with instruments
funded by ESA member states, and with special contributions from Denmark
and NASA (USA). The Planck-LFI project is developed by an International
Consortium led by Italy and involving Canada, Finland, Germany, Norway,
Spain, Switzerland, UK, USA.\\ 
The Italian contribution to Planck is
supported by the Italian Space Agency (ASI).\\ 
The Spanish participation is funded by Ministerio
de Ciencia e Innovacion through the projects
ESP2004-07067-C03-01 and AYA2007-68058-C03-02\\
The US Planck Project is supported by the NASA Science
Mission Directorate.\\
In Finland, the Planck project was supported by the Finnish
Funding Agency for Technology and Innovation (Tekes).

\newpage

\begin{appendix} 

\section{Radiometer output equation}

The output of a Back End detector can be analytically expressed as:

\begin{multline}
 V_{out} = a\cdot k \cdot\beta\cdot G_B \{\left[\left(G_{F1} + G_{F2} \right)\cdot\left(\frac{G_{F1}T_{nF1} + G_{F2}T_{nF2}}{2L_{WG}} \right)
 +\left( 1-\frac{1}{L_{WG}}\right)\cdot T^{eff}_{WG} + T_{nB}\right]\cdot(1-r) \\
 +\frac{\sqrt{G_{F1}G_{F2}}}{L_{WG}}\left( T_{sky} - T_{ref} \right)\cdot(1+r) \}
\label{full_out}
\end{multline}

where:
\begin{itemize}
\item $a$ is the square law detector constant
\item $\beta$ is the radiometer bandwidth
\item $k$ is the Boltzmann constant
\item $L_i$ are the RF losses of the different stages of the radiometer chain
\item $G_i$ and $T_{ni}$ are the FEM and BEM amplifiers gains and noise temperatures
\item $T_{sky}$ and $T_{ref}$ are the equivalent sky and ref antenna temperatures exiting the Front End.
\item $T^{eff}_{WG}$ is the effective waveguide temperature as integrated through its route.
\end{itemize}

More details can be found in \cite{LFI,seiffert}

\section{Thermal susceptibility detailed results}

In this section a summary of the transfer function measurements is given. 
Temperature dependance of LNAs gain and noise temperature are also estimated from the comparison between the measured and the analytically estimated transfer functions, in the hypoteses of perfectly symmetrical coupled amplifiers.\\

\begin{table}[!ht]
 \caption{RCA18 thermal susceptibility results}
 \centering
	\begin{tabular}{l c c c c}
\hline
Ch						& 		S-11				&				S-10				&				M-00			 &			M-01						\\
\hline
\\
$\frac{\partial G}{\partial T}$ (dB/K)			&  -0.05 $\pm$ 0.005			&			-0.05	$\pm$ 0.005		&			-0.05	$\pm$ 0.005	 &		-0.05 $\pm$ 0.005   \\
\\
$\frac{\partial Tn}{\partial T}$ (K/K)				& 	0.42 $\pm$ 0.02			&			0.38 $\pm$ 0.02		&			0.47 $\pm$ 0.02		 &		0.49 $\pm$ 0.02      \\
\\
\hline
\\
${\rm f_{meas}}$ (mK/K)				& -77 $\pm$ 3 & -68 $\pm$ 4 & -85 $\pm$ 5 & -91 $\pm$ 4       \\
\hline
 \end{tabular}
\end{table}

\begin{table}[!ht]
 \caption{RCA19 thermal susceptibility results}
 \centering
	\begin{tabular}{l c c c c}
\hline
Ch						& 		S-11				&				S-10				&				M-00			 &			M-01						\\
\hline
\\
$\frac{\partial G}{\partial T}$ (dB/K)			& -0.03	$\pm$ 0.005			&			-0.023 $\pm$ 0.005		&			 -0.05 $\pm$ 0.005	 &			-0.035 $\pm$ 0.005      \\
\\
$\frac{\partial Tn}{\partial T}$ (K/K)				& 0.37 $\pm$ 0.02				&			0.4 $\pm$ 0.02				&			 0.36 $\pm$ 0.02		 &			0.33  $\pm$ 0.02       \\
\\
\hline
${\rm f_{meas}}$ (mK/K)					& -111 $\pm$ 10 & -120 $\pm$ 9 & -94 $\pm$ 7 & -9 $\pm$ 8     \\
\hline
 \end{tabular}
\end{table}  
                                                                              
\begin{table}[!ht]
 \caption{RCA20 thermal susceptibility results}
 \centering
	\begin{tabular}{l c c c c}
\hline
Ch						& 		S-11				&				S-10				&				M-00			 &			M-01						\\
\hline
\\
$\frac{\partial G}{\partial T}$ (dB/K)			& -0.039 $\pm$ 0.005			&			-0.03	$\pm$ 0.005		&		-0.049 $\pm$ 0.005		 &			-0.0408 $\pm$ 0.005    \\     
\\
$\frac{\partial Tn}{\partial T}$ (K/K)				& 0.25 $\pm$ 0.02				&			0.3	$\pm$ 0.02			&			0.25 $\pm$ 0.02		 &			0.23 $\pm$ 0.02        \\           
\\
\hline                                                                              
${\rm f_{meas}}$ (mK/K)					& -58 $\pm$ 7 & -59 $\pm$ 7 & -66 $\pm$ 8 & -57 $\pm$ 8     \\
\hline
 \end{tabular}
\end{table}

\begin{table}[!ht]
 \caption{RCA21 thermal susceptibility results}
 \centering
	\begin{tabular}{l c c c c}
\hline
Ch						& 		S-11				&				S-10				&				M-00			 &			M-01						\\
\hline
\\
$\frac{\partial G}{\partial T}$ (dB/K)			& 	-0.2	$\pm$ 0.05		&			-0.07	$\pm$ 0.005		&			-0.07	$\pm$ 0.005	 &			-0.07 $\pm$ 0.005      \\        
\\
$\frac{\partial Tn}{\partial T}$ (K/K)				& 	0.3	$\pm$ 0.02			&			0.18 $\pm$ 0.02			&			0.15 $\pm$ 0.02		 &			0.15 $\pm$ 0.02        \\           
\\
\hline
${\rm f_{meas}}$ (mK/K)					& -9.3 $\pm$ 0.8 & -7.7 $\pm$ 0.9 -30.1 $\pm$ 0.9 & -18 $\pm$ 1     \\
\hline
 \end{tabular}
\end{table}  
                                                                              
\begin{table}[!ht]
 \caption{RCA22 thermal susceptibility results}
 \centering
	\begin{tabular}{l c c c c}
\hline
Ch						& 		S-11				&				S-10				&				M-00			 &			M-01						\\
\hline
\\
$\frac{\partial G}{\partial T}$ (dB/K)			& 	-0.13	$\pm$ 0.005		&			-0.175 $\pm$ 0.005		&			-0.213 $\pm$ 0.005	 &			-0.15 $\pm$ 0.005      \\    
\\
$\frac{\partial Tn}{\partial T}$ (K/K)				& 	0.1	$\pm$ 0.02			&			0.1	$\pm$ 0.02			&			0.1 $\pm$ 0.02			 &			0.1 $\pm$ 0.02        \\              
\\
\hline
${\rm f_{meas}}$ (mK/K)					& 60.4 $\pm$ 0.4 & 56.9 $\pm$ 0.8 & 56.7 $\pm$ 0.6 & 52.1 $\pm$ 0.7      \\
\hline
 \end{tabular}
\end{table}  
                                                                              
\begin{table}[!ht]
 \caption{RCA23 thermal susceptibility results}
 \centering
	\begin{tabular}{l c c c c}
\hline
Ch						& 		S-11				&				S-10				&				M-00			 &			M-01						\\
\hline
$\frac{\partial G}{\partial T}$ (dB/K)			& 	-0.05	$\pm$ 0.005		&			-0.05	$\pm$ 0.005		&			-0.03	$\pm$ 0.005	 &			-0.05 $\pm$ 0.005      \\         
\\
$\frac{\partial Tn}{\partial T}$ (K/K)				& 	0.16 $\pm$ 0.02			&			0.17 $\pm$ 0.02			&			0.1 $\pm$ 0.02	  	 &			0.16 $\pm$ 0.02       \\           
\\
\hline                                                                              
${\rm f_{meas}}$ (mK/K)					& -39 $\pm$ 3 & -41 $\pm$ 4 & -21 $\pm$ 2 & -44 $\pm$ 3     \\
\hline
 \end{tabular}
\end{table}  
                                                                              
\begin{table}[!ht]
 \caption{RCA24 thermal susceptibility results}
 \centering
	\begin{tabular}{l c c c c}
\hline
Ch						& 		M-00				&				M-01				&				S-11			 &			S-10						\\
\hline
\\
$\frac{\partial G}{\partial T}$ (dB/K)			& 	-0.08	$\pm$ 0.005		&			-0.063 $\pm$ 0.001		&			-0.08	$\pm$ 0.005	 &		-0.081 $\pm$ 0.001       \\     
\\
$\frac{\partial Tn}{\partial T}$ (K/K)				& 	0.4	$\pm$ 0.02 			&				0.41 $\pm$ 0.02		&				0.43 $\pm$ 0.02	 &			0.1 $\pm$ 0.02        \\           
\\
\hline                                                                              
${\rm f_{meas}}$ (mK/K)					& -12.1 $\pm$ 0.1 & -6.10 $\pm$ 0.08 & -9.6 $\pm$ 0.5 & -20.3 $\pm$ 0.7     \\
\hline
 \end{tabular}
\end{table}  
                                                                              
\begin{table}[!ht]
 \caption{RCA25 thermal susceptibility results}
 \centering
	\begin{tabular}{l c c c c}
\hline
Ch						& 		S-11				&				S-10				&				M-00			 &			M-01						\\
\hline
\\
$\frac{\partial G}{\partial T}$ (dB/K)			& 		-0.045 $\pm$ 0.005	&				-0.042 $\pm$ 0.005	&				-0.022 $\pm$ 0.005 &			-0.02 $\pm$ 0.005      \\   
\\
$\frac{\partial Tn}{\partial T}$ (K/K)				& 		0.08 $\pm$ 0.02		&				0.25 $\pm$ 0.02		&				0.12 $\pm$ 0.02	 &			0.1 $\pm$ 0.02        \\         
\\
\hline
${\rm f_{meas}}$ (mK/K)					& -22.1 $\pm$ 0.7 & -29 $\pm$ 2 & -15.1 $\pm$ 0.9  & -13 $\pm$ 0.5   \\
\hline
 \end{tabular}
\end{table}  
                                                                              
\begin{table}[!ht]
 \caption{RCA26 thermal susceptibility results}
 \centering
	\begin{tabular}{l c c c c}
\hline
Ch						& 		M-00				&				M-01				&				S-10			 &			S-11						\\
\hline
\\
$\frac{\partial G}{\partial T}$ (dB/K)			& 		-0.01	$\pm$ 0.005	&				-0.026 $\pm$ 0.005	&				-0.01	$\pm$ 0.005 &			-0.01 $\pm$ 0.005      \\
\\
$\frac{\partial Tn}{\partial T}$ (K/K)				& 		0.7	$\pm$ 0.02		&				0.7	$\pm$ 0.02		&				0.47 $\pm$ 0.02	 &			0.5 $\pm$ 0.02         \\
\\
\hline                                                                              
${\rm f_{meas}}$ (mK/K)					& -66 $\pm$ 2 & -64 $\pm$ 2 & -68 $\pm$ 1 & -65.6 $\pm$ 0.9    \\
\hline
 \end{tabular}
\end{table}  
                                                                              
\begin{table}[!ht]
 \caption{RCA27 thermal susceptibility results}
 \centering
	\begin{tabular}{l c c c c}
\hline
Ch						& 		M-00				&				M-01				&				S-11			 &			S-10						\\
\hline
\\
$\frac{\partial G}{\partial T}$ (dB/K)			& 		-0.055 $\pm$ 0.005	&				-0.05	$\pm$ 0.005	&				-0.01 $\pm$ 0.005	 &			-0.04 $\pm$ 0.005      \\       
\\
$\frac{\partial Tn}{\partial T}$ (K/K)				& 		0.81 $\pm$ 0.02		&				0.45 $\pm$ 0.02		&				0.34 $\pm$ 0.02	 &			0.58 $\pm$ 0.02        \\         
\\
\hline
${\rm f_{meas}}$ (mK/K)					& -16.8 $\pm$ 0.5 & -10.5 $\pm$ 0.5 & -31 $\pm$ 3 & -36 $\pm$ 2     \\
\hline
 \end{tabular}
\end{table}  
                                                                              
\begin{table}[!ht]
 \caption{RCA28 thermal susceptibility results}
 \centering
	\begin{tabular}{l c c c c}
\hline
Ch						& 		S-10				&				S-11				&				M-00			 &			M-01						\\
\hline
$\frac{\partial G}{\partial T}$ (dB/K)			& 		-0.13 $\pm$ 0.005		&			-0.14	$\pm$ 0.005		&			-0.03	$\pm$ 0.005	 &			-0.067 $\pm$ 0.005     \\                        
\\
$\frac{\partial Tn}{\partial T}$ (K/K)				& 		0.33 $\pm$ 0.02		&			0.1 $\pm$ 0.02				&			0.15 $\pm$ 0.02		 &			0.15 $\pm$ 0.02        \\                          
\\
\hline
${\rm f_{meas}}$ (mK/K)					& -11 $\pm$ 2 & -19 $\pm$ 2 & -2.7 $\pm$ 0.7 & 5.2 $\pm$ 0.8      \\
\hline
 \end{tabular}
\end{table}

\end{appendix}

\end{document}